\documentstyle[12pt,aaspp]{article}
\def\singlespace {\smallskipamount=3.75pt plus1pt minus1pt
                  \medskipamount=7.5pt plus2pt minus2pt
                  \bigskipamount=15pt plus4pt minus4pt
                  \normalbaselineskip=15pt plus0pt minus0pt
                  \normallineskip=1pt
                  \normallineskiplimit=0pt
                  \jot=3.75pt
                  {\def\smallskip {\vskip\smallskipamount}}
                  {\def\medskip   {\vskip\medskipamount}}
                  {\def\bigskip   {\vskip\bigskipamount}}
                  {\setbox\strutbox=\hbox{\vrule
                    height10.5pt depth4.5pt width 0pt}}
                  \parskip 7.5pt
                  \normalbaselines}
\def\middlespace {\smallskipamount=5.625pt plus1.5pt minus1.5pt
                  \medskipamount=11.25pt plus3pt minus3pt
                  \bigskipamount=22.5pt plus6pt minus6pt
                  \normalbaselineskip=22.5pt plus0pt minus0pt
                  \normallineskip=1pt
                  \normallineskiplimit=0pt
                  \jot=5.625pt
                  {\def\smallskip {\vskip\smallskipamount}}
                  {\def\medskip   {\vskip\medskipamount}}
                  {\def\bigskip   {\vskip\bigskipamount}}
                  {\setbox\strutbox=\hbox{\vrule
                    height15.75pt depth6.75pt width 0pt}}
                  \parskip 11.25pt
                  \normalbaselines}
\def\doublespace {\smallskipamount=7.5pt plus2pt minus2pt
                  \medskipamount=15pt plus4pt minus4pt
                  \bigskipamount=30pt plus8pt minus8pt
                  \normalbaselineskip=30pt plus0pt minus0pt
                  \normallineskip=2pt
                  \normallineskiplimit=0pt
                  \jot=7.5pt
                  {\def\smallskip {\vskip\smallskipamount}}
                  {\def\medskip   {\vskip\medskipamount}}
                  {\def\bigskip   {\vskip\bigskipamount}}
                  {\setbox\strutbox=\hbox{\vrule
                    height21.0pt depth9.0pt width 0pt}}
                  \parskip 15.0pt
                  \normalbaselines}

\def\etal   {{\sl et~al.}}
\def\lax    {${_<\atop^{\sim}}$}
\def\gax    {${_>\atop^{\sim}}$}
\def\mo     {{M$_{\odot}$}}

\def\kms    {~km~s$^{-1}$}

\def\kms    {~km~s$^{-1}$}

\def\myref#1  {\noindent \hangindent=24.0pt \hangafter=1 {#1} \par}
\def\bigref#1  {\noindent \hangindent=24.0pt \hangafter=2 {#1} \par}
\def\figs#1#2 {\item{#1} {#2} }

\begin{document}

%
%

\title{Observations of the X-ray Nova GRO~J0422+32:\\
II: Optical Spectra Approaching Quiescence.\altaffilmark{1}}

\vspace{0.5in}

\author{Michael R. Garcia}
\author{Paul J. Callanan}
\author{Jeffrey E.  McClintock}
\author{Ping Zhao}
\vspace{0.25in}

\affil{Center for Astrophysics, 60 Garden Street, Cambridge, MA
02138}

\affil{E-mail:  mgarcia,pcallanan,jmcclintock,pzhao, @cfa.harvard.edu}

\vspace{1.5in}
\centerline{
{\sl Submitted to ApJ:} \underline{1995 May 10}, Revised
\underline{1995 August 31};
}
\vspace{1.5in}
{\sl Subject Headings:}  accretion disks, stars:individual(GRO
J0422+32)

\altaffiltext{1}{Data reported herein was obtained with the Whipple
Observatory MMT, an instrument jointly operated by the Smithsonian
Astrophysical Observatory and the University of Arizona.}

\clearpage
\newpage

\title{Observations of the X-ray Nova GRO~J0422+32:\\
II: Optical Spectra Approaching Quiescence.\altaffilmark{1}}

\author{Michael R. Garcia}
\author{Paul J. Callanan}
\author{Jeffrey E.  McClintock}
\author{Ping Zhao}
\affil{Center for Astrophysics, 60 Garden Street, Cambridge, MA
02138}
\affil{E-mail:  mgarcia,pcallanan,jmcclintock,pzhao, @cfa.harvard.edu}
\altaffiltext{1}{Data reported herein was obtained with the Whipple
Observatory MMT, an instrument jointly operated by the Smithsonian
Astrophysical Observatory and the University of Arizona.}

\middlespace

\begin{abstract}
We present results obtained from a series of 5~\AA\ resolution spectra of the
X-ray Nova GRO~J0422+32 obtained in 1993~October, when the system was
approximately 2 magnitudes above quiescence, with ${\rm R \sim 19}$.  The data
were obtained in an effort to measure the orbital radial velocity
curve of the secondary, but detection of the narrow photospheric
absorption lines needed to do this proved elusive.  Instead we found
wide absorption bands reminiscent of M~star photospheric features.  The
parameters determined by fitting accretion disk line profiles (Smak
profiles) to the H$\alpha$ line are similar to those found in several
strong black-hole candidates. Measurements of the velocity of the
H$\alpha$ line are consistent with an orbital period of 5.1~hours and
a velocity semi-amplitude of the primary of $34 \pm 6$~\kms.  These
measurements, when combined with measurements of the velocity
semi-amplitude of the secondary made by others, indicate that the
mass ratio $q \sim 0.09$.  If the secondary follows the empirical
mass-radius relation found for CVs, the low $q$ implies
a primary mass of $M_x \sim 5.6$\mo, and a rather low (face-on)
inclination.  The H$\alpha$ EW is found to be modulated on
the orbital period with a phasing that implies a
partial eclipse of the disk by the secondary, but simultaneous R~band
photometry shows no evidence for such an eclipse.\\
\end{abstract}

\section{Introduction}

Low Mass X-ray Binaries (LMXBs) are systems where a neutron star or
black hole accretes material from a low mass (M$\leq$1M$_{\odot}$)
secondary (see Bhattacharya and van den Heuvel 1991).  X-ray novae (XRN)
form a subset of LMXBs whose properties are characterized by dramatic
increases in flux (factors of $\sim$10$^{5}$--10$^{7}$) at X-ray and
radio energies on time scales of days, with a subsequent decay to
quiescence on time scales of months.

While these transients are in outburst, their X-ray and optical
properties are very similar to those of the persistently bright LMXBs,
where the optical flux is dominated by the reprocessed emission from
the X-ray irradiated disk.  In quiescence, however, the disk often
fades to reveal the secondary star itself. This allows detailed
photometric and spectroscopic measurements of the secondary to be made
-- otherwise impossible in the outburst state, and in most of the
persistently bright LMXBs.

Studies of quiescent XRN are especially important for the following
reasons.  The frequency of black hole candidates among X-ray novae
appears to be remarkably high; 9 out of the 13 optically identified
systems are strong or probable black hole candidates (van Paradijs and
McClintock 1995).  Our understanding of the evolution of LMXBs (eg, the
progenitors, lifetimes, and numbers) will be limited until we
understand the nature of the transient sources.  This is because the
number of ``dormant'' transients may well exceed the number of
persistent LMXBs by one to two orders of magnitude (van den Heuvel
1992).

The detection of the photospheric lines of the secondary in quiescent
systems allows
measurement of the orbital radial velocity curve, leading to firm
lower limits on the mass of the accreting object.
These measurements indicate masses
\gax~3\mo\/ for the X-ray nova A0620-00 (McClintock and
Remillard 1986), V404 Cyg (Casares, Charles and Naylor 1992) Nova
Muscae 1991 (Remillard, McClintock and Bailyn 1992), Nova Ophiuci
1977 (Remillard \etal\/ 1995), GRO~J1655-40 (Bailyn \etal\/ 1995), and
GS2000+25
 (Casares, Charles and Marsh 1995).  Because the maximum mass for a stable
neutron star is $\sim 3$\mo\/ (Chitre and Hartle 1976) these six XRN
are regarded as strong black hole candidates.  The mass ratio $q$ can
be determined by combining the measurements of the orbital velocity of
the secondary with similar measurements of emission lines originating
in the accretion disk.  Models for formation of emission lines in
accretion disks depend on many parameters, among them the inclination,
inner and outer disk radii, and central object mass, and comparison to
observed profiles can help to constrain some of these parameters.


In Callanan \etal\/ 1995 (Paper~I) we discussed the discovery of
GRO~J0422+32, and presented the first results from our optical
observing campaign during the outburst and decline.  In this paper we
present the results of extensive spectroscopic observations obtained
$\sim 430$~days after outburst, when GRO~J0422+32 had decayed to ${\rm
R}\sim 19$ (see Figure~1).  Our continuing photometric observations
show that the source subsequently reached its quiescent level at day
$\sim 760$.  The average intensity during days 760--950 is ${\rm R} =
20.94 \pm 0.11$ (28~measurements) and ${\rm V}=22.35\pm 0.17$
(3~measurements).  We note that the error on the R~magnitude in
quiescence quoted here is computed from the observed scatter in the
data, and may be dominated by intrinsic variability in the source.
The outburst amplitude was $\Delta {\rm V} \sim 9.0$, which is the
largest yet seen for an XRN: typical outburst amplitudes are $\Delta
{\rm V} = 5-7$~magnitudes (van Paradijs and McClintock 1995).

\newpage

\section{Observations}

	By 1993~October~10, $\sim 14$ months after the initial
outburst, GRO~J0422+32 had reached ${\rm R} \sim 19$ (day 430 on
Figure~\ref{r.long.lc}), but was still at least $\sim 2$~magnitudes
brighter than its quiescent level.  This is surprising given that the
black hole candidates Nova Muscae 1991 (Remillard, McClintock and
Bailyn 1992), GS2000+25 (Chevalier and Ilovaisky 1990), and V404 Cyg
(Casares and Charles 1992) had reached quiescence within 14
months of the outburst.

	Spectra of GRO~J0422+32 were obtained with the SAO Whipple
Observatory MMT Red Channel Spectrograph on 1993~October~10 and 12 (UT), and
spectra of several bright standards were obtained on October~11 (UT) for
use in determining absolute radial velocities and spectral types.  The
spectra were obtained with a Loral $1200\times 800$~pixel CCD having
7~electrons read noise.  The combination of the grating and $1''$ slit
gave $\sim 5$~\AA~ FWHM spectral resolution and spectral coverage from
$\sim 5000$~\AA~through $\sim 6700$~\AA.  The seeing on October~10 and 12 was
predominantly $\sim 1''$ and the skies were mostly clear, but occasional
clouds and intervals of worse seeing ($\sim 2''$) compromised the
photometric accuracy of our spectra.  The cloud cover on October~11 was
sufficiently heavy that only bright standards were visible on the TV
guider; however GRO~J0422+32 was easily visible on the other two nights.
The slit was approximately maintained at the parallactic angle in
order to maximize throughput.  Twenty minute integrations of GRO~J0422+32
were bracketed by short exposures of a HeArNe lamp for wavelength
calibration purposes.  Data were reduced and analyzed using standard
IRAF routines.  Sixteen spectra were obtained on each night for a
total exposure time of 640~minutes on GRO~J0422+32.
Simultaneous Whipple Observatory 1.2~m
R~band photometry was obtained; the instrumentation
and data reduction are described in Callanan \etal\/ 1995.

\newpage

\section{Analysis and Results}
\subsection{Cross-Correlation Results}

	Our primary goals were to detect the expected photospheric
absorption lines due to the presumed late-type companion, and to
measure the Doppler shifts in those lines due to the expected orbital
motion of the star.  We attempted this by cross-correlating the
32~individual GRO~J0422+32 spectra against the spectra of late-type
dwarf templates, ranging from early K to mid~M, which were obtained
with the identical instrumental setup on Oct~11.  Before
cross-correlating, the regions around the 5577~\AA\ and 5890~\AA\
night-sky lines were replaced with a fit to the local continuum,
because these night sky lines are strong enough to render subtraction
in the individual 20~minute exposures problematic.  In addition, the
spectra were filtered in order to remove low-frequency features and in
order to match the instrumental resolution of 5~\AA.  The low
frequency filtering was achieved by subtraction of a $13^{\rm th}$
order cubic spline fit to the continuum, and removal of wavenumbers $<
6$ from the Fourier transform of the spectra (FT).  This effectively
removes all features on scales larger than $\sim 100$\AA, including
the instrumental response and possible TiO bands (see Section 3.2).
The high frequency filter removed wavenumbers $> 330$ from the FT,
which has the effect of smoothing the data on the intrinsic resolution
of 5\AA\/ ($\sim 3.5$ pixels per resolution element).  In order to
avoid the contaminating effects of the H-Balmer emission lines, only
the region between 5000~\AA\ and 6500~\AA\ was used for the
correlations.

We were not able to detect any narrow, late-type photospheric lines in
this data set, despite extensive efforts.  In order to quantify our
lack of detection, we added various amounts of a late-type stellar
spectrum (with NaD excised) to our individual GRO~J0422+32 spectra, and
computed cross-correlations.  In keeping with the short period, we
added dwarf (rather than giant) spectra.
When a M3V spectrum accounts for $\sim 10$\%,
a M0V for $\sim
20\%$, or a K3V for $\sim 30\%$ of the total light, the
correlations yield significant detections
with R (Tonry and Davis 1979) values in excess of 3.0.  In this way we
estimate that the fractional contribution of any late-type stellar
photosphere to the continuum light at ${\rm R} \sim 19$ is less than
$10\% \pm 5\%$,
$20\% \pm 5\%$, or
$30\% \pm 5\%$
for an assumed M3V, M0V or K3V companion, respectively.  \\

\subsection{Shape of the Continuum}

	Figure~\ref{oct.spec} shows the sum of the spectra obtained on
October~10~(bottom) and October~12~(top).  The most obvious features
are double peaked H$\alpha$ emission and interstellar NaD~5890~\AA~
absorption.  The NaD night sky line has not been excised because it is
accurately subtracted in the summed spectrum, while the 5577~\AA\/ has
once again been removed.  The NaD EW (1.2~\AA) is consistent with an
interstellar origin.  Less obvious in Figure~\ref{oct.spec} are
troughs in the continuum with minima at  $\sim
6200$~\AA, $\sim 5900$~\AA, and $\sim 5200$~\AA.  These troughs are
more apparent in Figure~\ref{oct.spec.bands}, which shows a scaled
plot of the October~10 GRO~J0422+32 spectrum and a spectrum of a
M3V~star.  The correspondence in the broad continuum features leads us
to suspect that something like an M~star photosphere is present in the
spectra, even though cross-correlations show that the narrow lines
expected from an M star are not detectable.  We considered the
possibility that these troughs might be an artifact of our data
acquisition or reduction procedure; however, spectra of the nova
V849~Oph ($R \sim 18$) which were taken on the same nights with the
identical instrumental setup and reduced in the same way do not show
these features.  In addition, spectra taken by other observers a few
months later also show evidence for bands indicative of a late-type
stellar photosphere (Bonnet-Bidaud and Mouchet 1994), although the
lower S/N and resolution of these spectra make the bands difficult to
detect.  The evidence for these bands was much weaker in our spectra
from October~12.  This may be because the S/N in the October~12 data
was slightly lower than on October~10 (see Figure~\ref{oct.spec}), or
it may be that the band strength is variable and the $\sim
0.3$~magnitude brightening on October~12 is related to their
weaknesses.  We therefore concentrate solely on the October~10 spectra
for the remainder of this section.

	We measured the EW of the strongest two features (at 5900\AA\/
and 6200\AA) and compared them to the EW in the M3V star spectrum.  We
choose to compare to an M3V star because it is a good match to the
spectral type of GRO~J0422+32 determined in quiescence, which is
M2$\pm 1$V (Garcia \etal\/ 1994, Filippenko \etal\/ 1995).  The
continuum points and band limits are defined by the M3V star, and are
5805\AA-6039\AA\/ and 6145\AA-6350\AA, as marked on Figure~3.  The EW
of these bands in GRO~J0422+32 are $\sim 10$\AA, while they are $\sim
60$\AA\/ in the M3V star.  Thus we see that $1/6$, or $\sim 17$\% of the
light may be due to an M3V star.

	An independent estimate of the amount of apparent M~star
photosphere present in the spectra can be made by creating synthetic
spectra consisting of the sum of an M3V star and an accretion disk,
and looking for the best match to the GRO~J0422+32 spectra.  This
method is sensitive to the slope of the continuum, whereas the
previous method is sensitive only to the depth of the bands.  We first
de-reddened the GRO~J0422+32 spectra for $A_V = 1.2$ (see
Callanan~\etal\/ 1995), and then sum variable amounts of an M3V star
and an $F_{\lambda} = F * \lambda^{\alpha}$ accretion disk spectrum.
We choose a power law for the accretion disk spectrum because it is a
sufficiently accurate representation of what has been seen in
quiescent XRN.  We measured the slope of the powerlaw from the
published data on three XRN in quiescence.  The results are shown in
Table~1, along with the slope measured for GRO~J0422+32 during the
decay by Shrader \etal\/ 1994.
We then assumed the slope for GRO~J0422+32 was equal to the mean of
the values in Table~1, but in order to determine the errors below we
allowed $\alpha$ to vary over the full range shown in the table.

We then searched for the minimum $\chi^2$ between the observed and
synthetic spectra w.r.t. the fraction ($F$) of accretion disk.  The
minimum $\chi^2$ was not found to be unity, in part because the
observed spectrum does not accurately match the narrow lines of the
template M~star.  In order to estimate an error range for our fit, we
re-scaled the error bars so that the best fit $\chi^2 = 1$ (see
Shahbaz \etal\/ 1994), and computed 68\% confidence intervals at
$\chi^2_{\rm min}+1$, (Avni 1976).  In the best fit synthetic
spectrum, the M3V~star contributes between $20^{+11}_{-14}\%$ at 5000\AA\/ and
$50^{+18}_{-28}\%$ at 6500\AA, where the error ranges include the
a contribution due to the range in slopes found in
Table~1.

Given the rather substantial errors associated with measuring the
contribution based on the slope of the continuum, we see that both
measurements are consistent with the assumption that an M3V star
photosphere contributed \lax20\% of the light in October~1993, when
$R \sim 19$.  This is marginally consistent with our lack of a result
from the cross-correlations, as we would need $\sim 20$\% of the light to
be due to a late-type stellar photosphere in order to obtain reliable
cross-correlation results. \\

\subsection{H$\alpha$ profiles}

The double peaked H$\alpha$ profiles shown in Figure~\ref{oct.spec}
are similar to those seen in cataclysmic variables, where they are
believed to arise because of Doppler shifts in an accretion disk
around the compact object (Smak 1981, Horne \& Marsh 1986).  In order
to test this hypothesis for GRO~J0422+32 we have fit model profiles
appropriate for optically thin, flat, Keplerian disks.  The model and
fitting methods are described in detail by Smak 1981 and Orosz \etal\/
1994.  We restricted ourselves to models in which $\alpha = 1.5$
($\alpha$ is the exponent of the radially dependent emission
function), as this value is generally seen in dwarf nova disks in
quiescence (Horne 1993), and has also been found in XRN disks in
quiescence (Johnson, Kulkarni and Oke 1989, Orosz \etal\/ 1994).  If
we use other values for $\alpha$ the results of our model fits might
be different (see Figure 4 of Johnson, Kulkarni and Oke 1989), but by
restricting ourselves to $\alpha = 1.5$ we allow direct comparison to
previous work, in particular Orosz \etal\/ 1994.

As the steady state Smak model which we fit to the data predicts a
symmetrical emission line profile, we begin by selecting 17 of our 32
spectra from October~10 and 12 such that the sum has red and blue
peaks of equal amplitude.  A sum of all 32~spectra produces a profile
in which the blue peak dominates (Figure~\ref{oct.spec}).  This
departure from symmetry might be explained by phase or time variable
emission from an accretion disk hot spot seen in CVs and XRN (Johnson
\etal\/ 1989), or it might be due to poorly understood deviations from
optically thin, Keplerian models which have previously been noted in
XRN (Orosz \etal\/ 1994, Haswell and Shafter 1990).  We have not
attempted to correct the average H$\alpha$ profile for the possible
effects due to the motion of the compact primary or the H$\alpha$
absorption line expected from the secondary, as both of these effects
are expected to be small in GRO~J0422+32 and are typically ignored (ie,
Johnson, Kulkarni and Oke 1989, Orosz \etal\/ 1994).  In any case,
fits to the simple, symmetrical Smak models will allow comparison of
the GRO~J0422+32 line profile to those seen in other XRN and CVs
(Orosz \etal\/ 1994, Williams 1983.)

A first-order polynomial was fit to the 6350~\AA\ -- 6750~\AA\ region
(excluding H$\alpha$) in order to subtract the continuum from the
summed spectrum, and a chi-squared minimizing method was used to
determine the best fit model parameters.  The best fit parameters and
68\% confidence intervals that we found are: $r_1 = 0.14\pm 0.01$,
$\lambda_0 = 6564.0 \pm 0.1~{\rm \AA}$, and $v_d = 442 \pm 10$\kms,
were $r_1$ is the ratio of the radii of the inner and outer edges of
the H$\alpha$ emitting part of the disk, $\lambda_0$ is the central
wavelength of the profile, and $v_d$ is the velocity at the outer edge
of the disk (which is very nearly equal to 1/2 the separation of the
peaks, see Smak 1981).  A comparison of the data and best fit profile
can be seen in Figure~\ref{smak}. \\

\subsection{H$\alpha$ Radial Velocity}

	Because the H$\alpha$ line forms in the accretion disk, one
might expect that the line velocity reflects the motion of the compact
object.  Measurement of this velocity should then determine the
orbital period and mass function, $f(m)= PK_x^3/2\pi G = (M_c {\rm
sin}(i))^3 / (M_x + M_c)^2$, where $P$ is the orbital period, $K_x$ is
the semi-amplitude of the orbital velocity variation of the primary
(and of the H$\alpha$ line by our assumption), $M_c$ is the mass of
the companion, $M_x$ the mass of the compact star, and $i$ is the
orbital inclination.  The phase and mid-point ($\gamma$ velocity) of
the H$\alpha$ velocity curve relative to that of the secondary provide
checks on the validity of the assumption.  Studies of CVs (eg, Young,
Schneider and Shectman 1981, Stover 1981) and XRN (Haswell and Shafter
1990 hereafter HS90, Orosz \etal\/ 1994, Marsh, Robinson, and Wood
1994, hereafter MRW94) show that $K_x$ determined from H$\alpha$
emission lines is indicative of the velocity of the primary, even
though the phase and $\gamma$ of the H$\alpha$ velocity curve are
sometimes slightly offset from that expected.

However, the complex and variable shape of the H$\alpha$ line makes it
difficult to measure a velocity that reliably traces the motion of the
compact object.  Several methods have been developed to help alleviate
this problem, among them ``double Gaussian fitting'' (Schneider and
Young 1980, Shafter 1983, Marsh \etal\/ 1994) and Smak profile fitting
(Orosz \etal\/ 1994).  We have employed a variant of these methods.  A
single Gaussian is fit to the outer (red and blue) portion of the
H$\alpha$ line, excluding the central part of the line which can be
affected by bright spot emission, optical depth effects, and other
effects which may not reflect the motion of the compact object.  In
practice this outer portion is determined to be that portion in which
the second derivative of the flux vs wavelength is $>0$ as one
approaches the line center. Because this is the section of the line in
which the profile is steepest (the derivative is highest), its
location is very sensitive to changes in the line velocity.  We tested
our method on data obtained for XRN Muscae 1991 by Orosz and
collaborators in February~1993, and we find values of $K_x$ and $T_0$
(but not $\gamma$) that are consistent with those derived by the
double Gaussian fitting and phase resolved Smak profile fitting
methods (Orosz \etal\/ 1994).

	In order to improve the S/N and search for radial velocity
variations the individual 20~minute GRO~J0422+32 spectra were grouped
into 6 equal phase bins (with 4-7 spectra in each bin) based on the
suggested period of $P=5.0944 \pm 0.0017$~hours, (Chevalier and
Ilovaisky 1994b; we chose this period because it has the smallest
error among the many consistent periods reported to date).  For each
profile a $\chi^2$ minimizing method was used to determine the central
wavelength, width, and amplitude of a fitted Gaussian and linear
continuum model.  The errors in the fitted parameters were determined
by the rms to the fit.  A sine curve with period fixed at 5.0944~hours
was fitted to the velocities, which gave an amplitude $K_x = 34 \pm
6$~\kms and $\gamma = 142 \pm 4$ (see Figure~\ref{ha.vel}A).  For
comparison, Filippenko, Matheson and Ho (1995) find $K_x =42\pm 2$\kms
and $\gamma = 26.0 \pm 1.6$ from recent observations in quiescence
with the Keck telescope.  We adopt the phase convention of Orosz
\etal\/ 1994 and Johnson \etal\/ 1989, in which $T_0$ marks the
closest approach of the compact object to the observer, and find $T_0
={\rm HJD}~ 2,449,270.978\pm 0.004$.  \\

\subsection{H$\alpha$ EW and Continuum Variations}

	Previous studies of interacting binaries have shown that the
relative phasing of variations in emission line velocities, emission
line EW, and wide-band photometry has the potential to reveal much
about the geometry of a binary system (eg, Pringle and Wade 1985,
Mauche 1990).	In order to measure the EW of the H$\alpha$ emission
line, we fitted a low-order polynomial to the region around the
emission line in each of the 32 spectra, in order to determine the
continuum level.  Figure~\ref{ha.vel}B shows the resulting
H$\alpha$ EW vs the phase we determined above.  We find
that the EW is modulated by a factor of $\sim 2$ on the orbital
period, varying between 20~\AA\ and 40~\AA.  There is a broad minimum
in the EW around phase 0.5.

	We obtained 3~nights of R~band photometry with the SAO 1.2~m
telescope during October~1993.  Two of these nights were simultaneous
with the MMT spectra described herein, and are plotted vs phase in
Figure~\ref{ha.vel}C.  These data are also shown in
Callanan \etal\/ 1995, Figure~3(d) [but note that they are labeled via
MST dates, ie, October~9 and October~11].  The first seven
measurements on October~12 are significantly ($\sim 0.2$~magnitudes)
higher than the rest, indicating a flare-like event from GRO~J0422+32.
We excluded these points from the analysis below.  The solid line is
the average magnitude in each of 12 phase bins.  This average varies by
12\%, and the mean rms in each of the 12 bins is 5\%.
There appears to be a broad maximum between phases 0.5
and 1.0, with a slight indication of a local maximum around phase 0.5.

%
%

\section{Discussion}

	The apparently periodic modulation of the H$\alpha$ EW and
velocity reported herein agree with  the suggested
$\sim 5.1$~hour orbital period for GRO~J0422+32 (Chevalier and
Ilovaisky 1994b), which has recently been confirmed by spectroscopic
observations in quiescence (Orosz and Bailyn 1994, 1995; Filippenko
\etal\/ 1995, Casares \etal\/ 1995).

	The lack of a detection of the secondary from our
cross-correlation efforts indicates that the secondary contributed
only a small fraction of the light during October~1993.  The fact that
the system faded another 2~magnitudes since October~1993 by itself
sets an upper limit of 16\% to the contribution of the secondary in
October~1993.  The non-detection is consistent with tests of the
sensitivity of the cross-correlations to detecting the secondary in the
individual 20~minute spectra, which show that we would need a
contribution of $\sim 20\%$ for reliable detection (Section~3.2).
Somewhat surprisingly the secondary does seem to be discernable in the
grand sum spectrum, in that wide (but weak) M-star bands are seen in
the spectrum.  The measured strength of the bands may be somewhat larger
than the 16\% allowed contribution, however the errors in the
measurement are large, and given the tendency to overestimate the
strength of weak absorption features in noisy spectra it seems very
likely that the bands are due to the M2V secondary.

	The accretion disk parameters (and $1\sigma$ errors) derived
using the Smak profile fits for GRO~J0422+32 are compared to those
derived for the XRN A0620-00 and X-ray Nova Muscae~1991 (Orosz \etal\/
1994) in Table~2.  We see that the $r_1$ values derived for A0620-00
and GRO~J0422+32 are very similar, and are a factor of two larger than
that found for Nova Muscae.  Interpreted in the context of the Smak
model, this indicates that the ratio of the inner to outer radii of
the H$\alpha$ emitting part of the disk is $\sim 7$ in A0620-00 and
GRO~J0422+32, but $\sim 14$ in Nova Muscae.

	We can see in Figure~\ref{ha.vel} that the H$\alpha$ EW
decreases by $\sim 50\%$ at phase 0.5, about the same time that the
R~band photometry increases by $\sim 12\%$.  The relative amplitude of
the EW and R~band variations shows that only a small part of the the
H$\alpha$ EW variations can be due to variations in the continuum.
Casares \etal\/ (1995) also find a $\sim 50$\% orbital modulation in
the H$\alpha$ EW of GRO~J0422+32 in data taken in quiescence.  This
variation is in marked contrast to that seen in A0620-00 (MRW94) and
V404 Cyg (Casares and Charles 1992, Casares \etal\/ 1993), in which
the H$\alpha$ EW modulation is largely explained by modulation in the
underlying continuum.  The H$\alpha$ velocities imply that the
secondary is closest to the observer when the EW is smallest, which
suggests that the EW variations may be due to an eclipse of the disk
by the secondary.  However, this is untenable for a simple azimuthally
symmetric disk.  The problem is that any eclipse would have to cover
$\sim 50\%$ of the H$\alpha$ emitting region while at the same time
the overall R~band continuum (which must be due almost entirely to the
disk) peaks.  It is perhaps more likely that the EW variations are due
to variations in the projected geometry and integrated emissivity of
the disk itself, for example due to an azimuthally varying disk
thickness.

We searched for, and found, orbital radial velocity variations in the
H$\alpha$ emission line.  Such variations often measure the motion of
the compact object in SXT (Orosz \etal\/ 1994, HS90, MRW94).  In the
particular case of GRO~J0422+32, we can check to see if this velocity
variation is consistent with the compact object by comparing the
$\gamma$ velocity and phase to those measured for the secondary.

We find a central velocity of $\gamma = 142 \pm 4$\kms\/ for the
H$\alpha$ line in GRO~J0422+32, which is not consistent with velocity
of the secondary of $\gamma = 10 \pm 4$\kms (Filippenko \etal\/ 1995).
However, the $\gamma$ velocities derived from SXT H$\alpha$ studies
are somewhat problematic.  The values derived depend strongly on the
details of the function used to fit the line (ie, HS90, Orosz \etal\/
1994).  It is therefore not surprising that different methods lead to
different results.  In the best studies system, A0620-00, Orosz \etal\/
find $\gamma = 78 \pm 3$ via Smak profile fitting, but $\gamma = 1.5
\pm 0.8$ via double Gaussian fitting.  Even when consistent methods
are used, the H$\alpha$ $\gamma$ in A0620-00 appears to change from
year to year, perhaps because of variations in the shape of the
H$\alpha$ profile (using double Gaussian fitting, HS90 find $\gamma =
28\pm 6$ in Dec 89/Jan 90, Orosz \etal\/ find $\gamma = 1.5 \pm 0.8$ in
Jan 91/Feb 91, and MRW94 find $\gamma = 0 \pm 5$ in Dec 91/Jan 92).
One is lead to the conclusion that H$\alpha$~$\gamma$
velocities are not robust measures of the binary systemic velocity,
and that in the case of GRO~J0422+32 our measurement of $\gamma =142
\pm 4$\kms\/ argues neither strongly for nor against the assumption that
the H$\alpha$ velocity variations ($K_x$) are representative of the
compact object.  However, even given the problematic nature of the
H$\alpha$ $\gamma$ velocities, it appears that H$\alpha$ $K_x$
velocities are accurate, in that they yield results which agree
with those determined by independent methods (Orosz \etal\/ 1994, MRW94).

The orbital period of GRO~J0422+32 is not presently known with
sufficient accuracy to compare the phase of the secondary radial
velocity curve measured in quiescence (Filippenko \etal\/ 1995,
Casares \etal\/ 1995) to the H$\alpha$ curve we measured in October
1993.  However, the large H$\alpha$ EW variations, seen once per
orbital cycle, might provide an accurate fiducial mark.  If one
assumes this variation is stable with respect to the orbital phase,
then we can use the EW curve to compare our H$\alpha$ velocities to
those measured for the secondary.  Comparing our Figure~5~A/B to Figure~1
of Casares \etal\/ (1995), we see that the peak of the H$\alpha$ EW
modulation occurs when the H$\alpha$ velocity crosses from blue to
red, and the secondary velocity crosses from red to blue.  This is the
phasing expected if the H$\alpha$ velocities track the compact object.

By combining our measurement of the velocity semi-amplitude of the
primary ($K_x = 34 \pm 6$~\kms) with the recently determined
velocity semi-amplitude of the secondary ($K_c = 380 \pm
6$~\kms, Filippenko \etal\/ 1995) we can determine the mass
ratio $q = K_x/K_c = 0.089\pm 0.016$.  This mass ratio, while
extreme, is similar to that found in other XRN and reinforces the
idea that XRN as a class may have mass ratios extreme enough to excite
SU~UMa type superhumps in outburst (Callanan and Charles 1991, Bailyn 1992).
Indeed,  superhumps during outburst have been reported for
GRO~J0422+32 (Kato, Mineshige and Hirata 1995).

The mass function sets a lower limit to the mass of the compact object
of $ M_x = 1.21 \pm 0.04$\mo\/ (Filippenko \etal\/ 1995).  When we use
our estimate of the mass ratio (and the definition of the mass
function) we find the lower limit is raised to $ M_x > 1.43 \pm
0.09~sin^{-3}(i)$\mo.  The corresponding lower limit to the mass of
the secondary is $M_c > 0.12 \pm 0.02~sin^{-3}(i)$\mo.

A reasonable upper limit on the mass of the secondary might be derived
from the empirical mass/period relation found for secondaries in CVs
by Patterson (1984).  We feel this is an upper limit because studies
of Cen~X-4, for example,  show that the secondary in this X-ray binary is
substantially undermassive (Remillard and McClintock 1990).
For the 5.1~hour period of GRO~J0422+32, Patterson's
relation predicts $M_c = 0.5$\mo.  Given the $q$~found above, this
translates into an upper limit for the mass of the compact object of
$ M_x < 5.6\pm 1.4$\mo, and (through the mass function) a lower
limit to the inclination of $35^o$.  The lack of optical
and x-ray eclipses sets an upper limit to the inclination of $i <
80^o$ at the observed mass ratio (Chanan, Middleditch, and Nelson
1976).  The rather modest orbital modulation of the lightcurve
compared to that seen in other XRN favors the lower end of the
allowed range in $i$, and therefore the higher end of the allowed
range in $M_x$.  We note that the $q$ and $M_x$ we find are consistent
with those determined from the observations of superhumps during outburst
by Kato \etal\/ (1995).
However detailed modeling of the lightcurve (eg
Haswell \etal\/ 1993) will be
necessary to further constrain $i$, $q$, and $M_x$.

We thank the staff at the MMTO for their outstanding level of support,
and the multitude of observers at the WO 1.2~m telescope who helped
collect the extensive light curve of GRO~J0422+32.  This work was
partially supported by NASA contract NAS8-39073, grant NAGW-4296, and
an Oxford University Visitors Grant.  We thank Drs. Filippenko,
Casares, and Orosz for providing results prior to publication, and we
thank the referee for many helpful comments.


\newpage

\begin{center}
Table 1: \\
\vskip 10pt
\begin{tabular}{|c|c|c|l|}\hline
XRN	& Wavelength Range &	$\alpha$	& Reference	\\\hline
A0620-00 & 3000\AA - 10000\AA & -1.76	& Oke 1977 \\
XN Muscae 1991 & 5000\AA - 6500\AA & -1.88 & Orosz \etal\/ 1995 \\
V404 Cyg	& 4400\AA - 7000\AA & -1.63 & Casares \etal\/ 1993 \\
GRO~J0422+32	& 4000\AA - 7000\AA & -1.80 & Shrader \etal\/ 1994 \\\hline
\end{tabular}
\end{center}

\begin{center}
Table 2: \\
\vskip 10pt
\begin{tabular}{|l|ccc|}\hline
parameter       & A0620-00      & Nova Muscae   & J0422+32 \\\hline

$r_1$           & $0.15\pm 0.01$&$0.07\pm 0.01$ & $0.14 \pm 0.01$ \\
$v_d$           & $550\pm 10$   & $450 \pm 10$  & $442 \pm 10$ \\
\hline
\end{tabular}
\end{center}

\newpage

{\bf References} \\

\myref{Avni, Y., 1976 ApJ 210 642}


\myref{Bailyn, C.D., Orosz, J.A., McClintock, J.E., and Remillard, R.A., 1995,
s
ubmitted to NATURE}

\myref{Bailyn, C.D., 1992 ApJ 391, 298}

\myref{Bonnet-Bidaud, J.M., and Mouchet, M., 1994, submitted to A\&A}

\myref{Callanan, P.J., Garcia, M.R., McClintock, J.E., Zhao, P., 1995
ApJ in press}

\myref{Callanan, P.J., and Charles, P.A., 1991 MNRAS 249, 573}

\myref{Casares, J. Charles, P.A., and Marsh, T., 1995, submitted to MNRAS.}

\myref{Casares, J. \etal, 1995, accepted for publication in MNRAS}

\myref{Casares, J. Charles, P.A., and Naylor, T., Pavlenko, E.P, 1993
MNRAS 265, 834}

\myref{Casares, J. Charles, P.A., 1992 MNRAS 255, 7}

\myref{Casares, J. Charles, P.A., and Naylor, T., 1992 NATURE 355,
614}

\myref{Chanan, G.A., Middleditch, J., and Nelson, J.E., 1976 ApJ
208, 512}

\myref{Chitre, D.M., and Hartle, J.B.,  1976, ApJ 207, 592}

\myref{Chevalier, C. and Ilovaisky, S.A.,  1994b, IAUC 6118}

\myref{Chevalier, C. and Ilovaisky, S.A.,  1990 A\&A 238, 163}




\myref{Filippenko, A.V., Matheson, T., and Ho, L.C., 1995 ApJ, submitted}

\myref{Garcia, M.R., Callanan, P.J., McClintock, J.E., and Zhao, P.,
1994 BAAS 25, 1381}

\myref{Haswell, C.A.,  Robinson, E.L., Horne, K., Stiening, R.F., and
Abbott, T.M.C., 1993 Ap.J 411, 801}

\myref{Haswell, C.A.,  and Shafter, A.W.,  1990 ApJ 359, L47}

\myref{Horne, K., 1993, Proceedings NATO ARW ``Theory of Accretion
Disks II, Munich, Germany 22-25 March 1993}

\myref{Horne, K. and Marsh, T.R., 1986 MNRAS 218, 761}

\myref{Johnson, H.M., Kulkarni, S.R., and Oke, J.B. 1989, ApJ
345, 492}

\myref{Kato, T., Mineshige, S., Hirata, R., 1995, PASJ 47, 31.}

\myref{McClintock, J.E., and Remillard, R.A., 1986 ApJ 308, 110}


\myref{Marsh, T.R., Robinson, E.L., and Wood, J.H., 1994,
MNRAS, 266, 137}

\myref{Mauche, C. W., 1990, Accretion Powered Compact Binaries,
(Cambridge: Cambridge University Press)}

\myref{Oke, J.B., 1977 ApJ 217, 181}

\myref{Orosz, J., and Bailyn, C., 1994, IAUC 6103}

\myref{Orosz, J.A., Bailyn, C.D., Remillard, R.A., McClintock, J.E.,
and Foltz, C.B.,  1994, ApJ 436, 848} 

\myref{Orosz, J.A., Bailyn, C.D., McClintock, J.E., and Remillard,
R.A., 1995, ApJ (submitted)} 

\myref{Orosz, J., and Bailyn, C., 1995, submitted to ApJ}

\myref{Patterson, J., 1984 ApJ Suppl 54, 443}


\myref{Pringle, J.E., and Wade, R.A,  1985, Interacting Binary Stars,
(Cambridge, Cambridge University Press)}

\myref{Remillard, R.A.,  Orosz, J.A., McClintock, J.E.,  and Bailyn,
C.D.,  1995, ApJ, submitted}

\myref{Remillard, R.A.,  McClintock, J.E.,  and Bailyn, C.D.,  1992
ApJ 399, L145} 

\myref{Remillard, R.A., and  McClintock, J.E.,  1990 ApJ 350, 386}

\myref{Ritter, H, and Kolb, U., 1995, to appear in ``X-ray binaries'',
W.H.G. Lewin, J. van Paradijs, E.P.J. van den Heuvel (Eds.),
Cambridge University Press (in press)}

\myref{Schneider, D.P. and Young P.J. 1980 ApJ 238, 946}

\myref{Shafter, A.W., 1983 ApJ 267, 222}


\myref{Shahbaz, T., Ringwald, F.A., Bunn, J.C., Naylor, T., Charles,
P.A., and Casares, J. 1994 MNRAS 271, L10}

\myref{Shrader, C.R., Wagner, R.M., Hjellming, R.M, Han, X.H., and
Starrfield, S.G., 1994 ApJ 434, 698}

\myref{Smak, J. 1981 Acta Astr. 31.4, 395}

\myref{Stover, R.J., 1981 ApJ 248, 684}

\myref{Tonry, J. and Davis, M. 1979 AJ 84, 1511}

\myref{Williams, G. 1983 ApJ Suppl, 53, 523}

\myref{Young, P., Schneider, D.P., and Shectman, S.A., 1981 ApJ 245,
1035}

\myref{van Paradijs, J. and McClintock, J.E., 1995, to appear in
``X-ray binaries'', W.H.G. Lewin, J. van Paradijs, E.P.J. van den
Heuvel (Eds.), Cambridge University Press (in press)}

\myref{van den Heuvel, E.P.J., 1992, Proc. Sat. Symp. Nr. 3, Int. Space Yr.
Conf., Munchen, (ESA ISY-3, July 1992), ESTEC, Noordwijk, p29}

\newpage
{\bf FIGURE CAPTIONS} \\

{\bf Figure 1}: The long term light curve of GRO~J0422+32 in the
R~band, beginning a few days after discovery in 1992~August and
continuing for more than two years.  The spectroscopic observations
reported here took place during days 430 and 432, in between the two
mini-outbursts (at days $\sim 370$ and $\sim 500$) which where were detected
during the decay.  The relative constancy of the data after day $\sim
760$ indicates that the system has now reached its quiescent magnitude
of ${\rm R} = 20.94 \pm 0.11$. An earlier version of this lightcurve
appeared in Callanan \etal\ 1995.\\

{\bf Figure 2:} The grand sum of the spectra obtained on
1993~October~10 (bottom) and October~12 (top).  The y-axis is flux in
units of ergs~cm${\rm ^{-2}sec^{-1}\AA^{-1}}$, with an offset of
0.2~added to the October~12 spectrum in order to clarify the plot.
Simultaneous photometry from the 1.2~m telescope indicated that the
mean brightness was slightly lower on October~10 and higher on
October~12, with $R=19.2$ and $R=18.9$, respectively.  The most
prominent feature in the spectra is the double peaked H$\alpha$
emission line. These spectra have not been de-redenned.  The EW of the
NaD line (1.2~\AA) is consistent with an interstellar origin.\\

{\bf Figure 3}: Summed spectrum of GRO~J0422+32 from 1993~October~10
compared to a M3V spectrum obtained as a cross-correlation template.
The GRO~J0422+32 spectrum has been de-reddened by $A_v=1.2$.
The straight lines indicate continuum points used to measure EW of
absorption bands.  \\

{\bf Figure 4:}
The summed H$\alpha$ profile of GRO~J0422+32 and best fitting Smak
model.  The Smak model is appropriate for optically thin, Keplerian
accretion disks.\\

{\bf Figure 5:} Panel A shows the wavelength of the center of the
H$\alpha$ emission line as observed at the telescope and as determined
from Gaussian fits to the steepest section of the line, assuming a
5.0944~hour orbital period, plotted vs phase.  The best fitting sine
curve of $K_x = 34 \pm 6$~\kms and heliocentric $\gamma = 142 \pm
4$~\kms\/ is superposed.  The time of closest approach of the primary is
$T_0 = {\rm HJD} 2,449,270.978 \pm 0.004$.  Panel B shows the EW of
H$\alpha$ vs the phase we have determined from the H$\alpha$
velocities.  Panel C shows the R magnitudes from the SAO 1.2~m
telescope, measured simultaneously with the H$\alpha$ EW.  The circles
are from October~10, the triangles from October~12, and the crosses
are the ``flare'' data (see text) from the beginning of October~12
offset by 0.2 magnitudes.  The solid line is the mean flux in each of
12~phase bins, and single error bar of 0.05~mag, representing the
mean rms of the binned data, is shown. \\

\clearpage
\newpage

\begin{figure}[h]
\hbox{
}
\caption{}\label{r.long.lc}
\end{figure}
\newpage

\begin{figure}[h]
\hbox{
\vspace{-1.5in}
\hspace{-1.0in}
}
\caption{}\label{oct.spec}
\end{figure}

\begin{figure}[h]
\hbox{
}
\caption{}\label{oct.spec.bands}
\end{figure}

\begin{figure}[h]
\caption{}\label{smak}
\end{figure}

\begin{figure}[h]
\caption{}\label{ha.vel}
\end{figure}

\begin{figure}[h]
\vspace{-0.5in}
\hbox{
\hspace{0.5in}
}
\caption{}\label{all.phase.photo}
\end{figure}

\begin{figure}[h]
\hbox{
\hspace{0.5in}
}
\caption{}\label{fwhm.vs.i}
\end{figure}

\clearpage

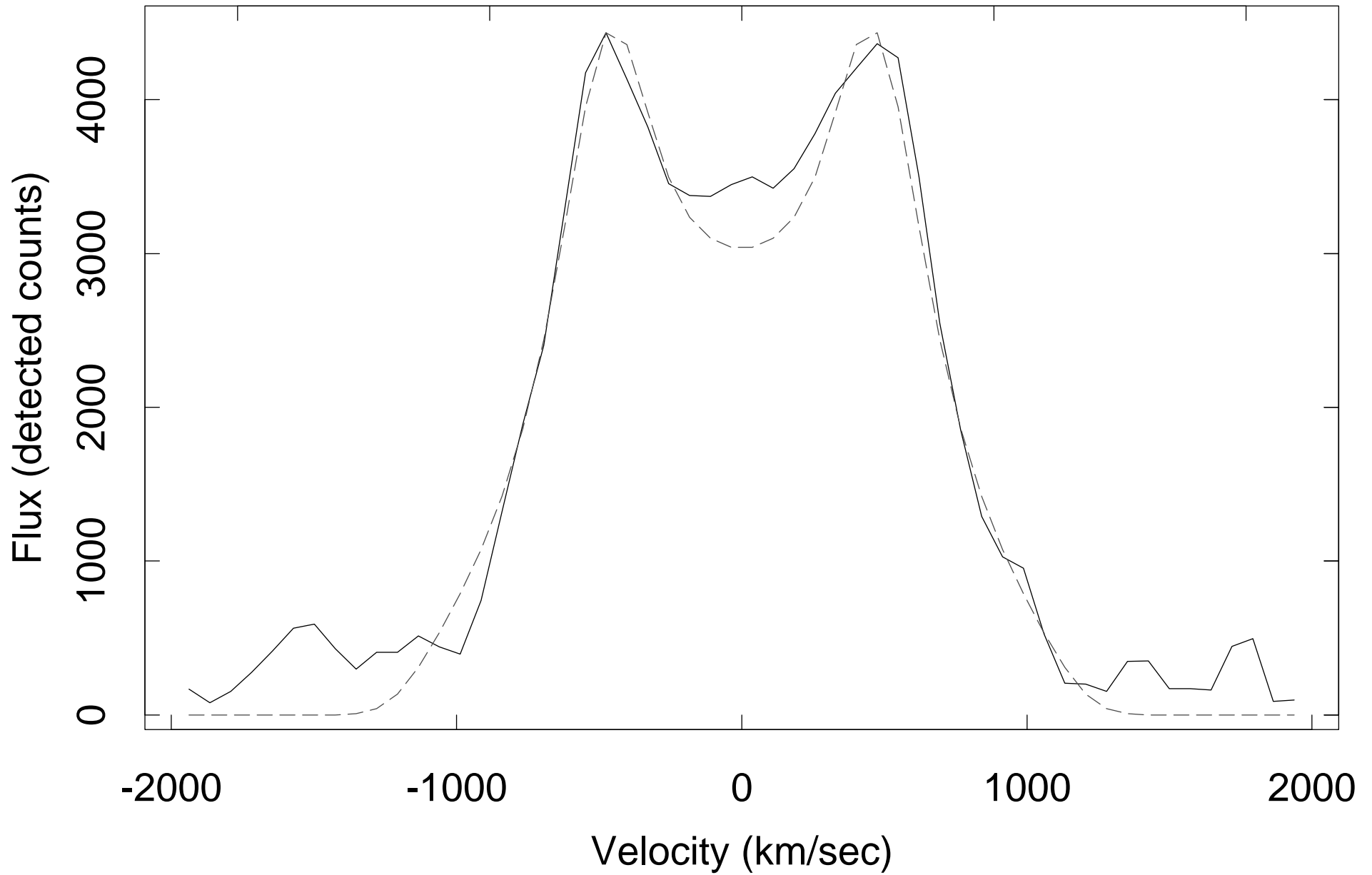
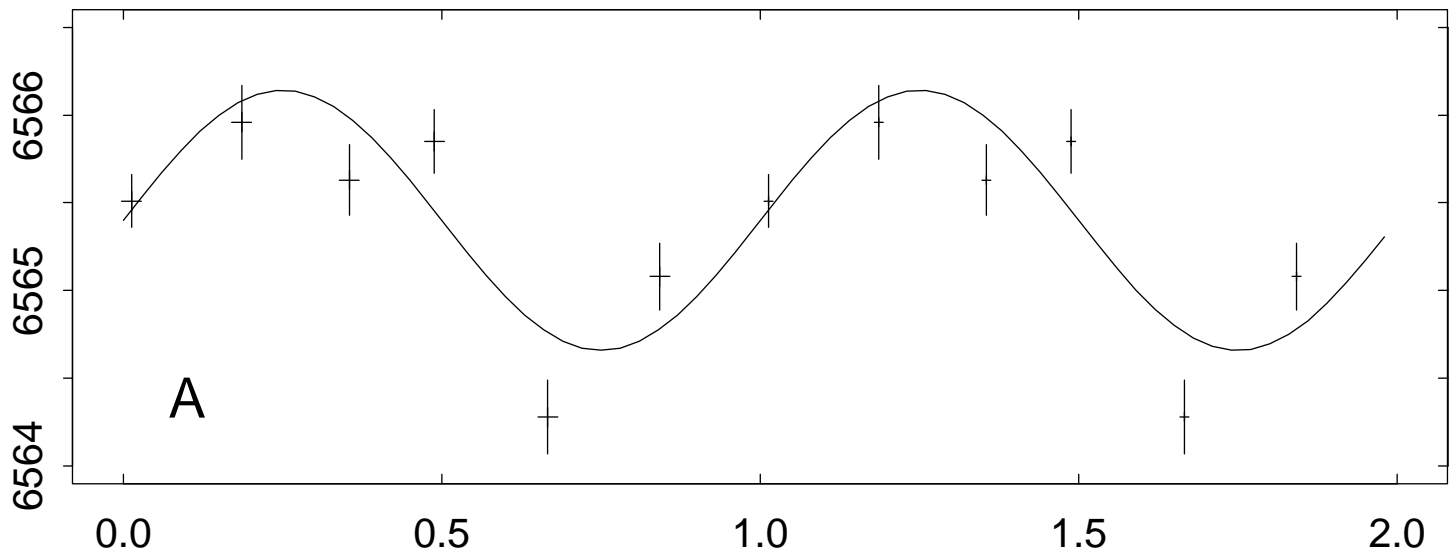
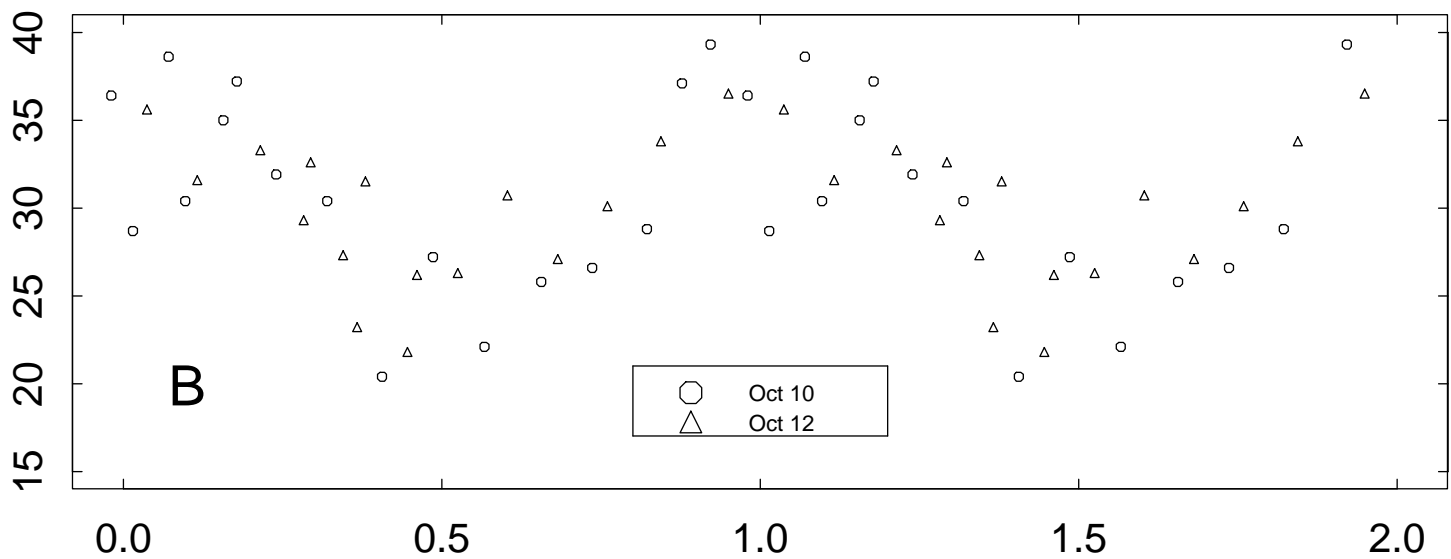
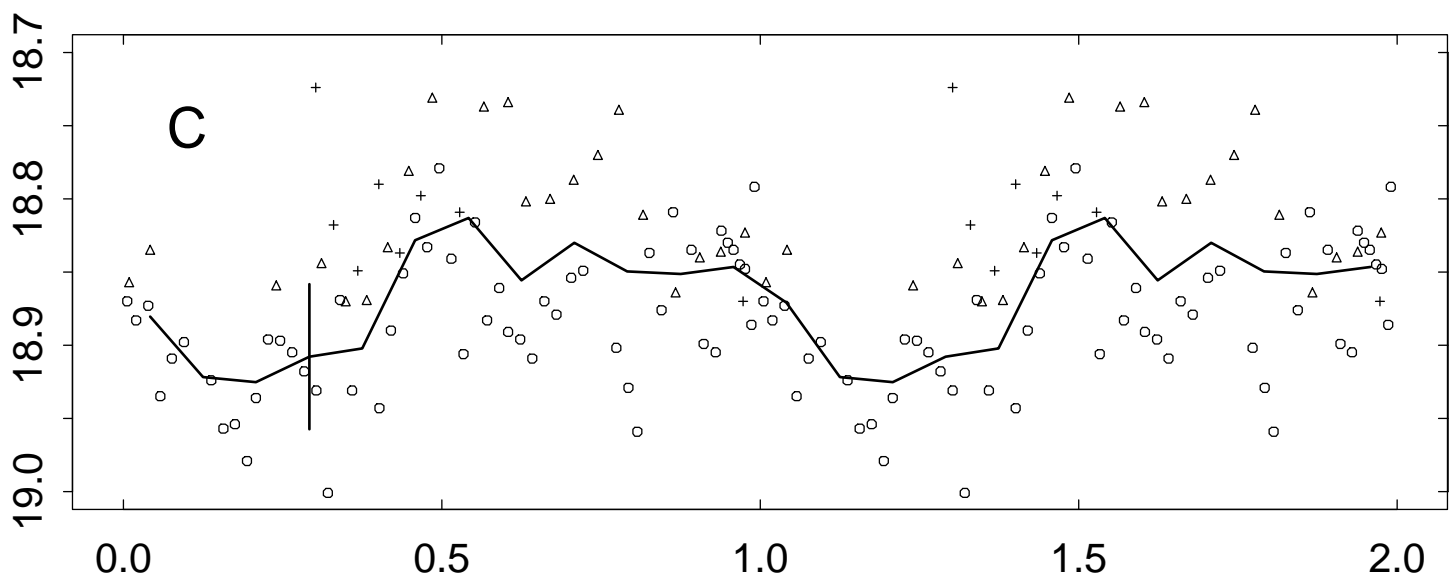
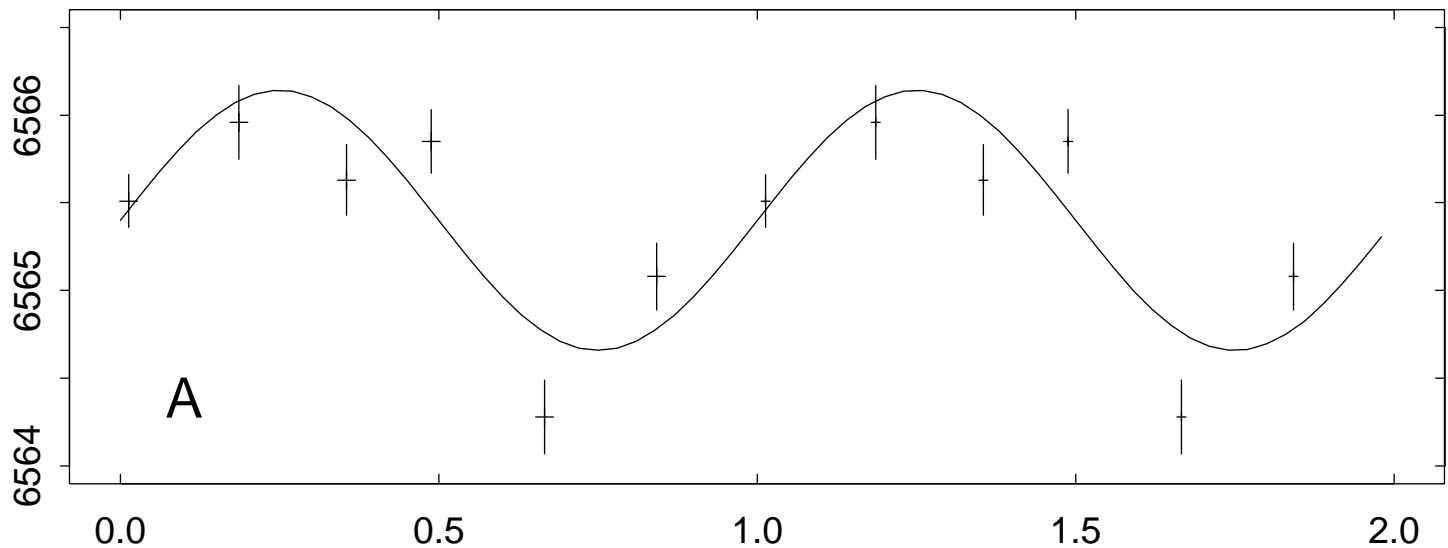
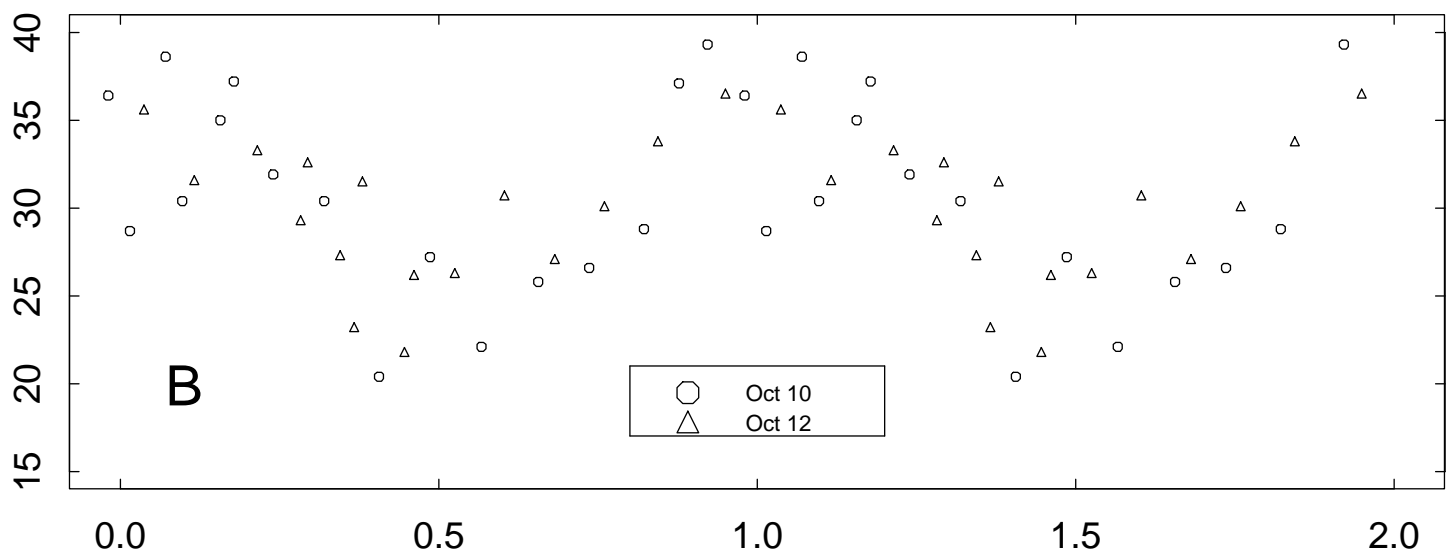
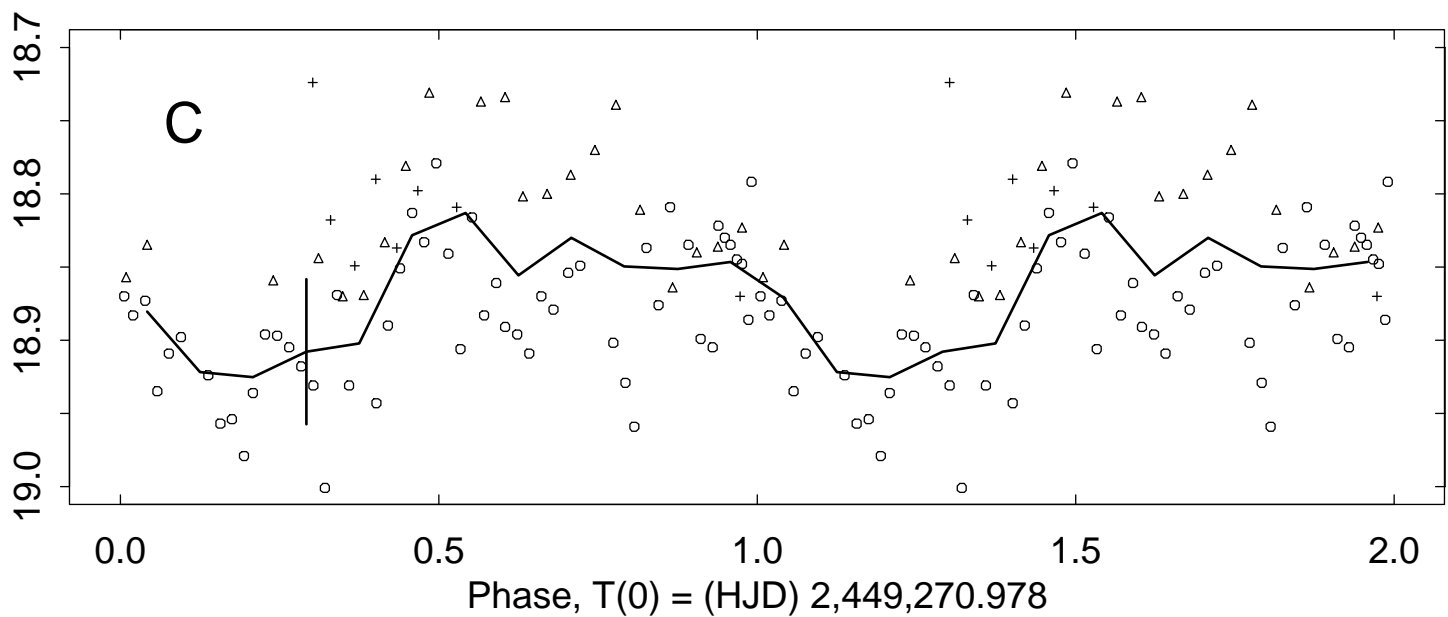
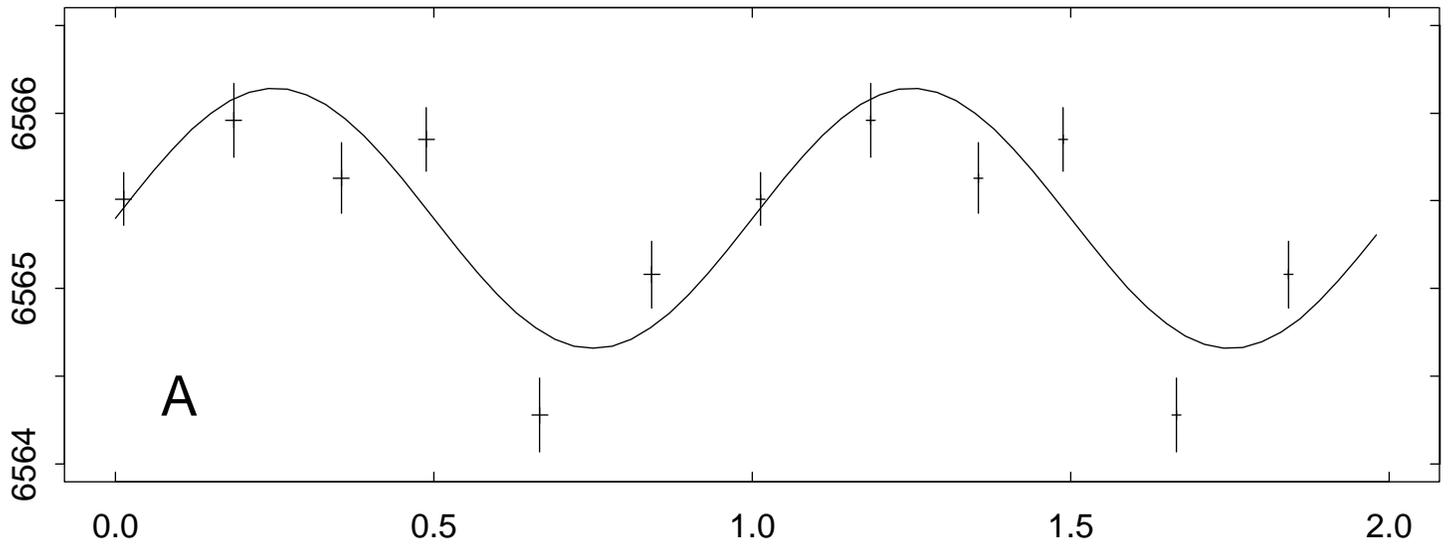
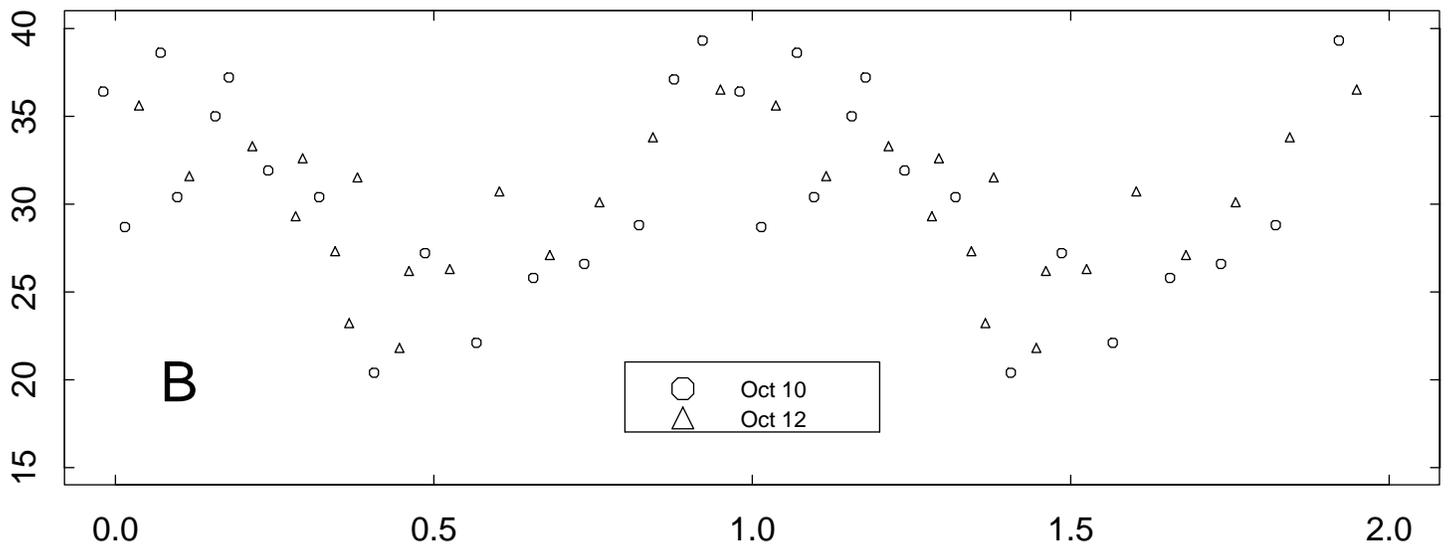
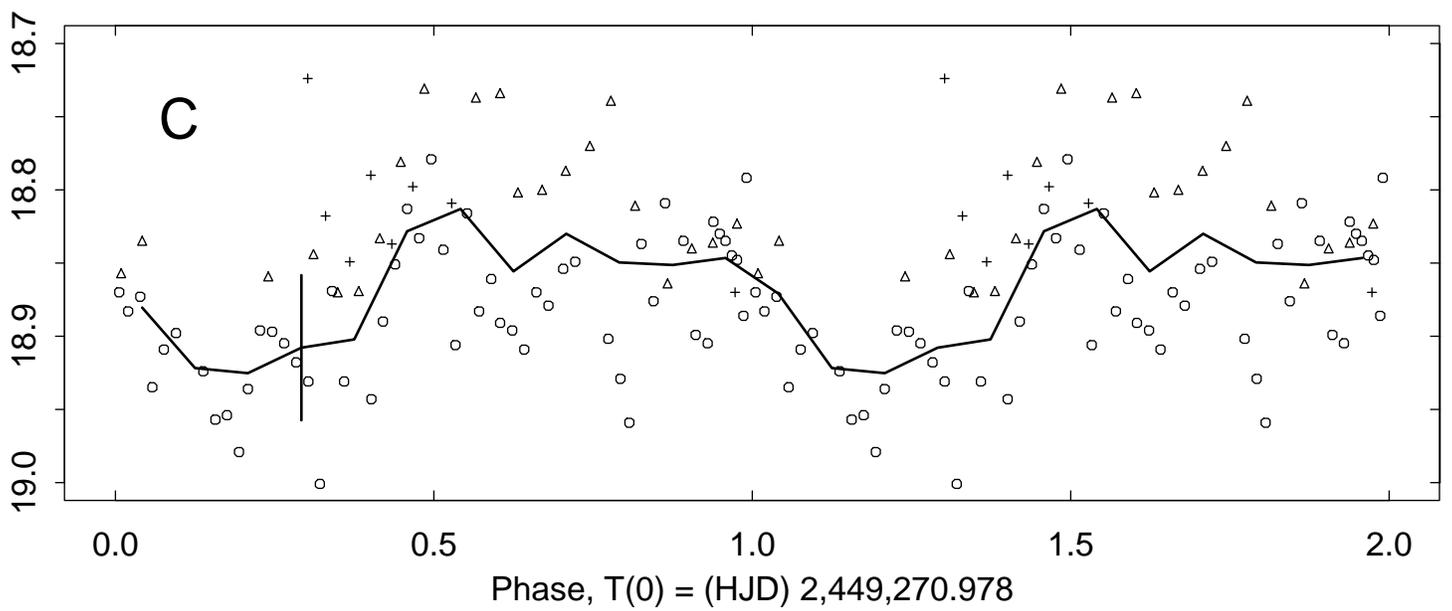
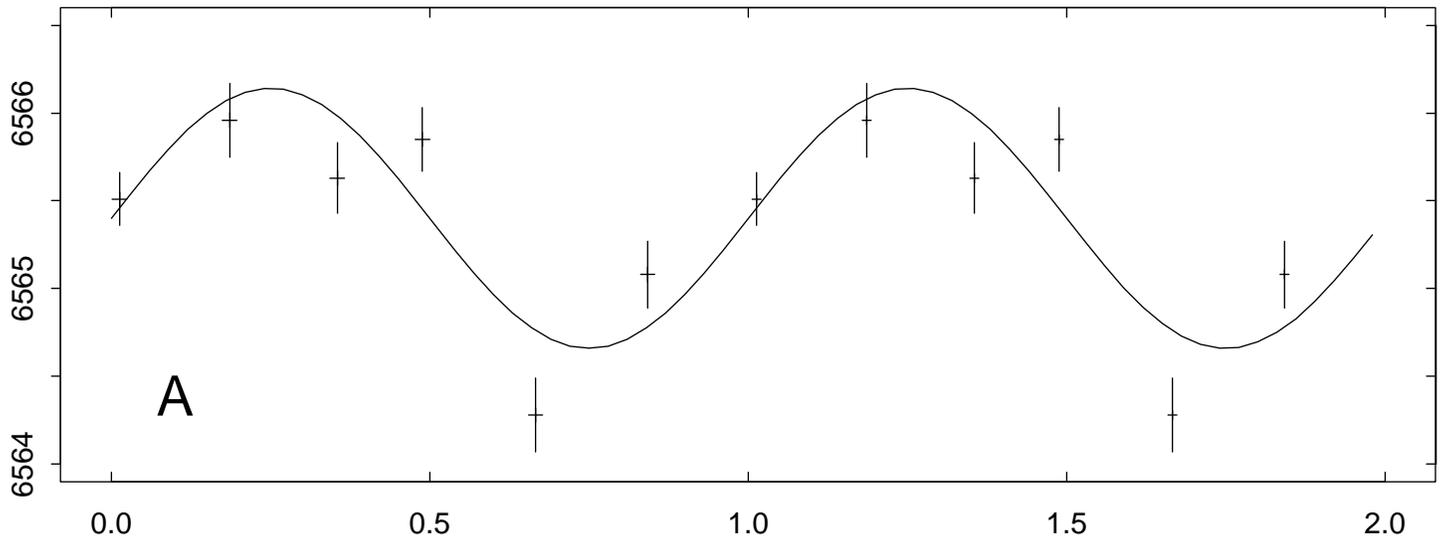
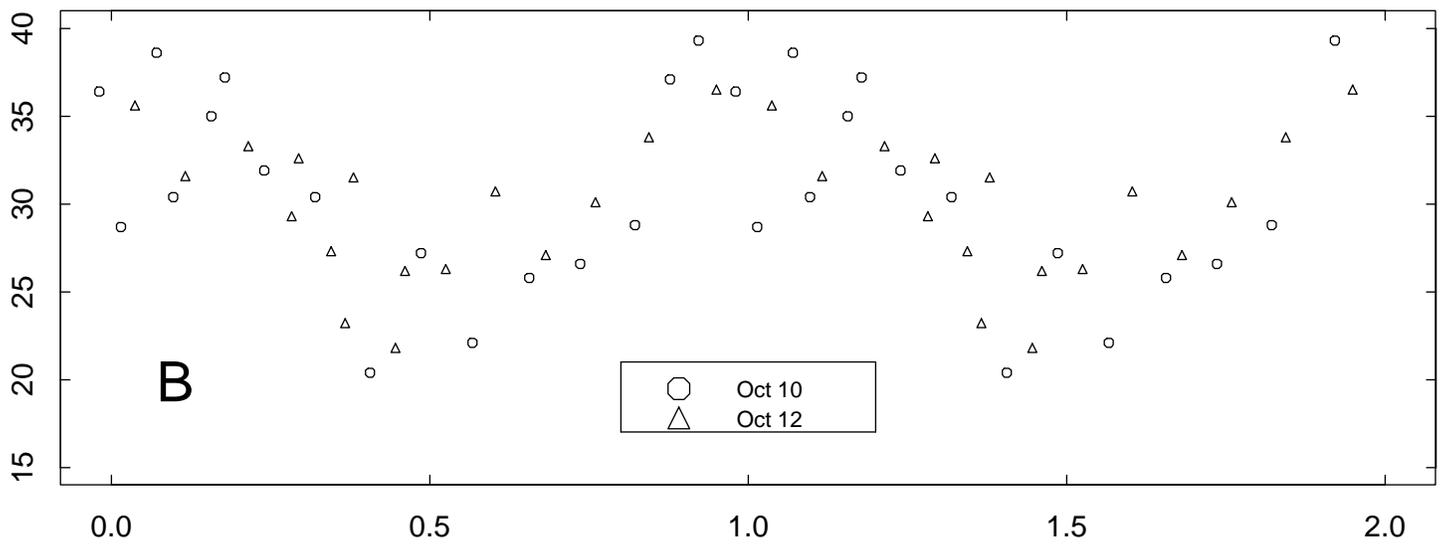
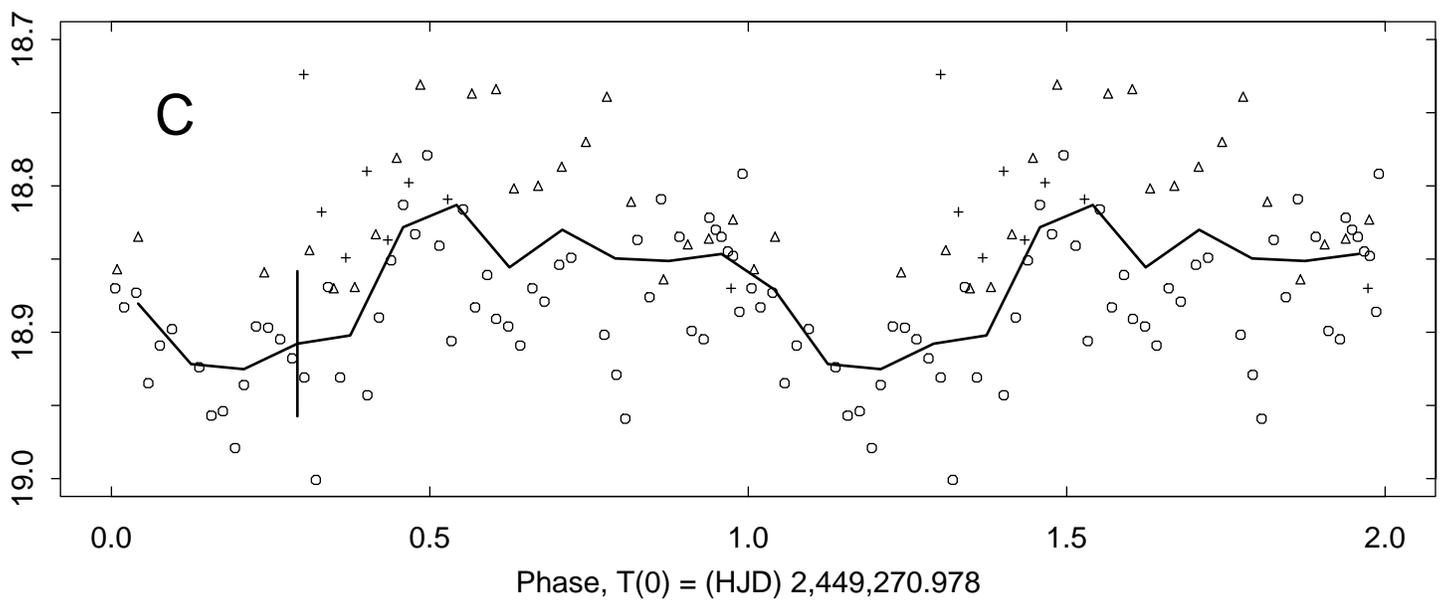
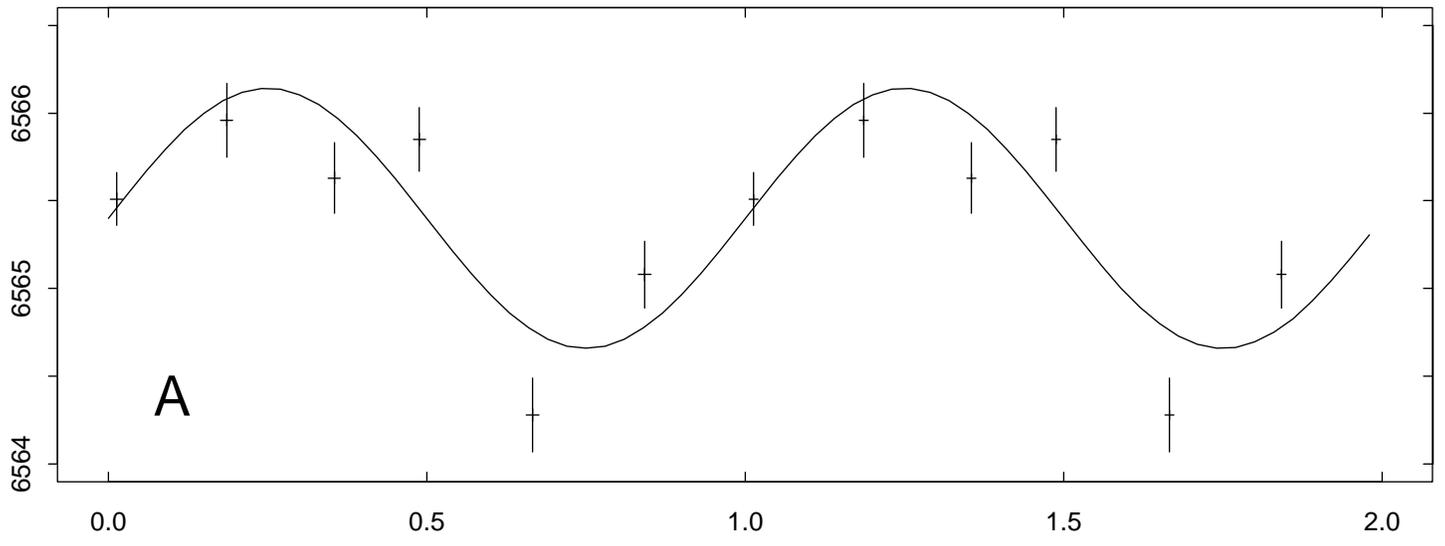
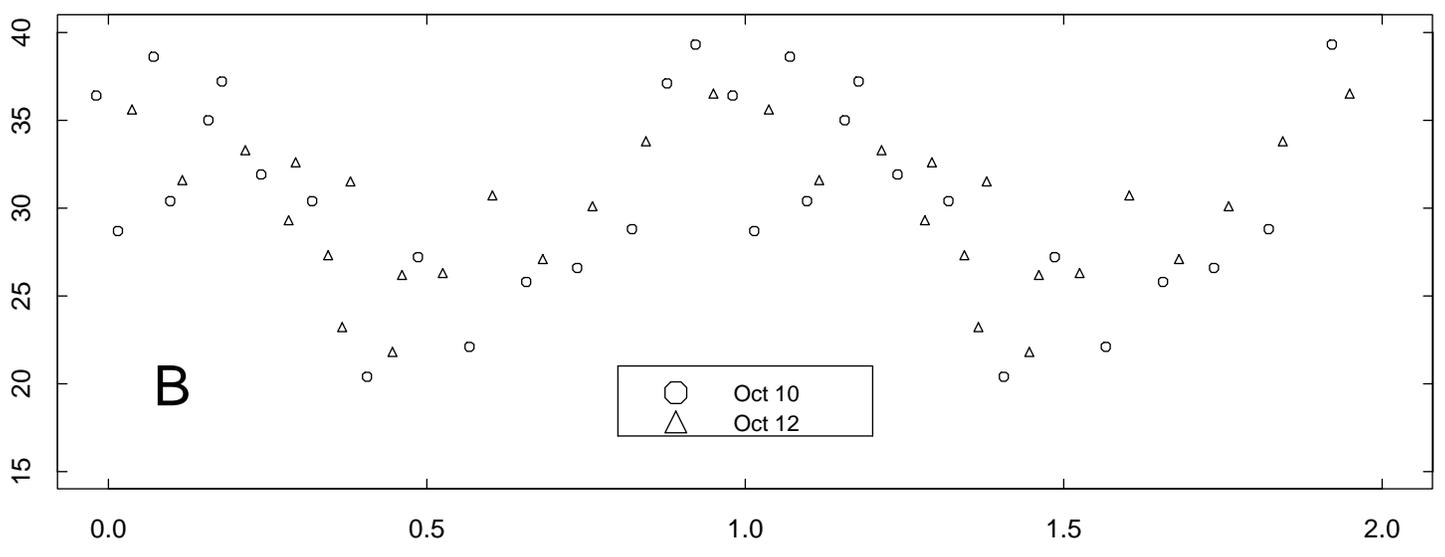
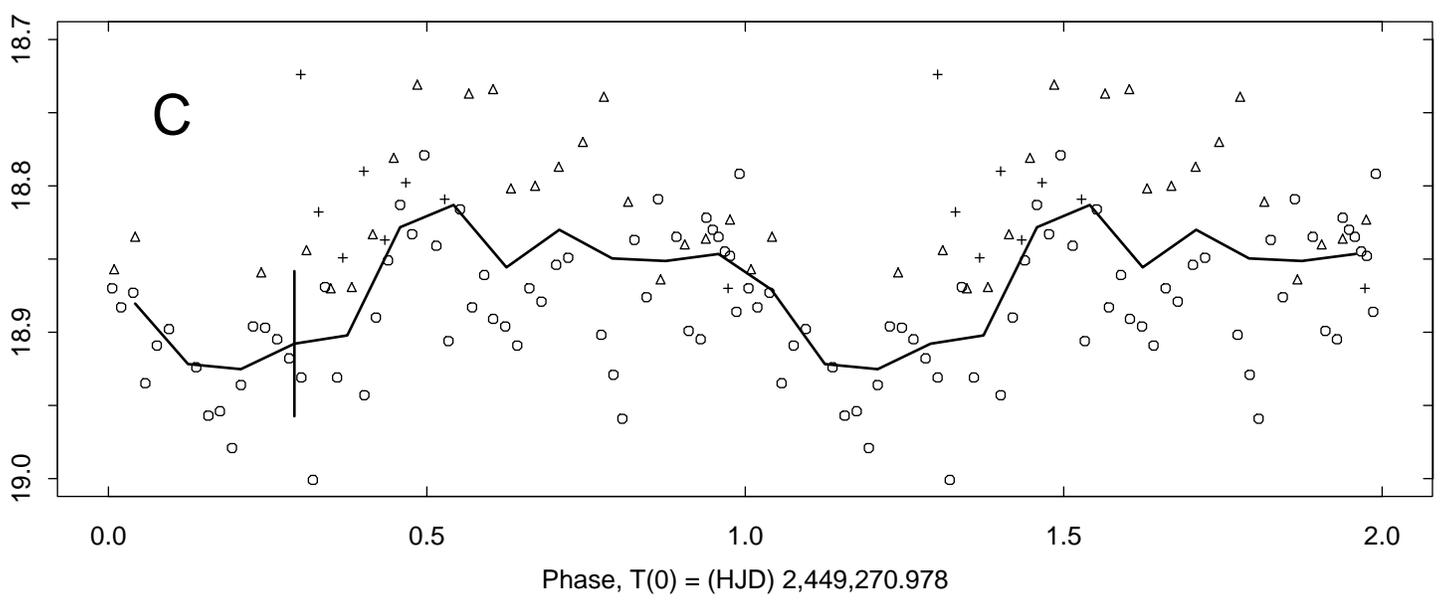
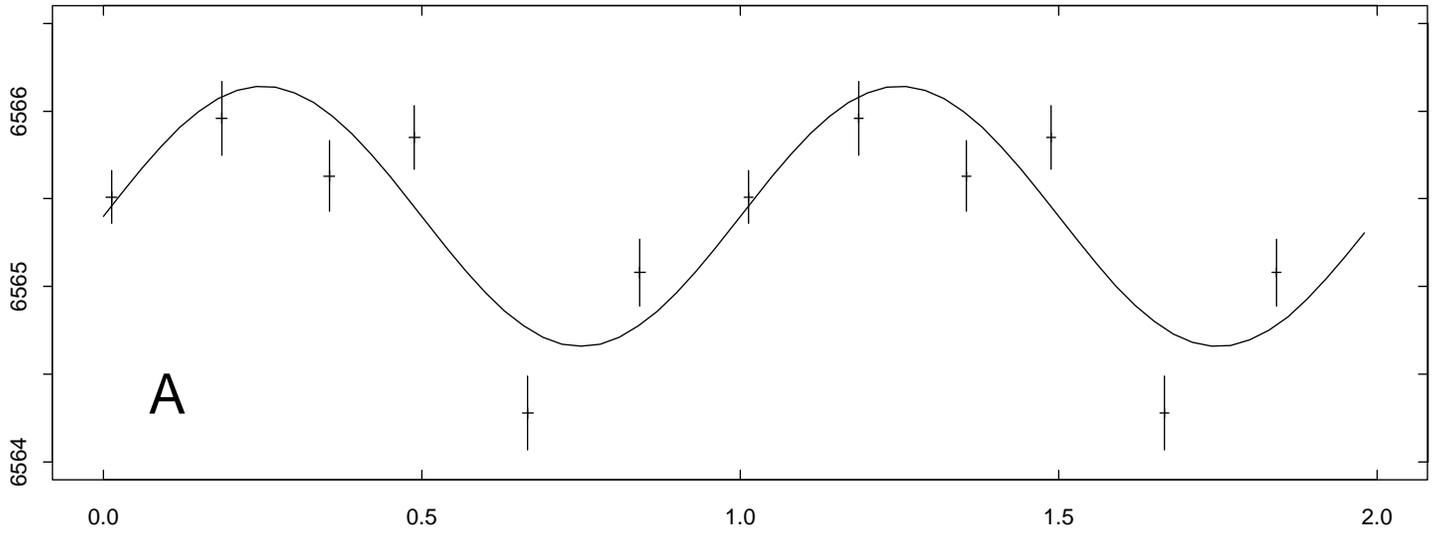
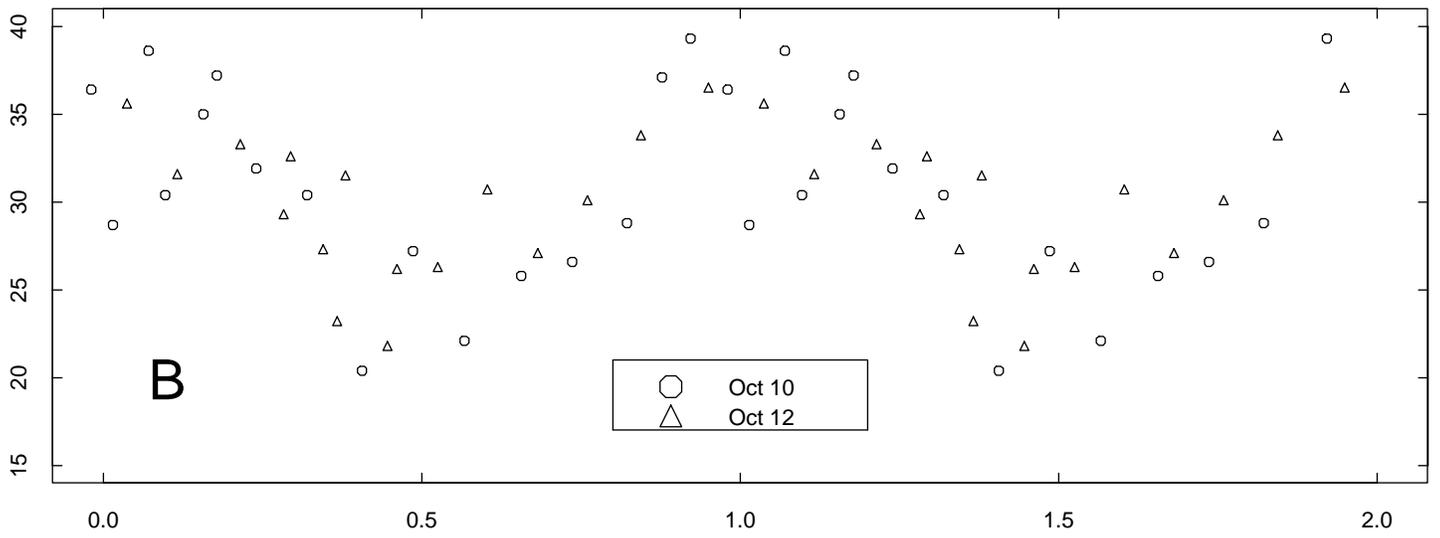
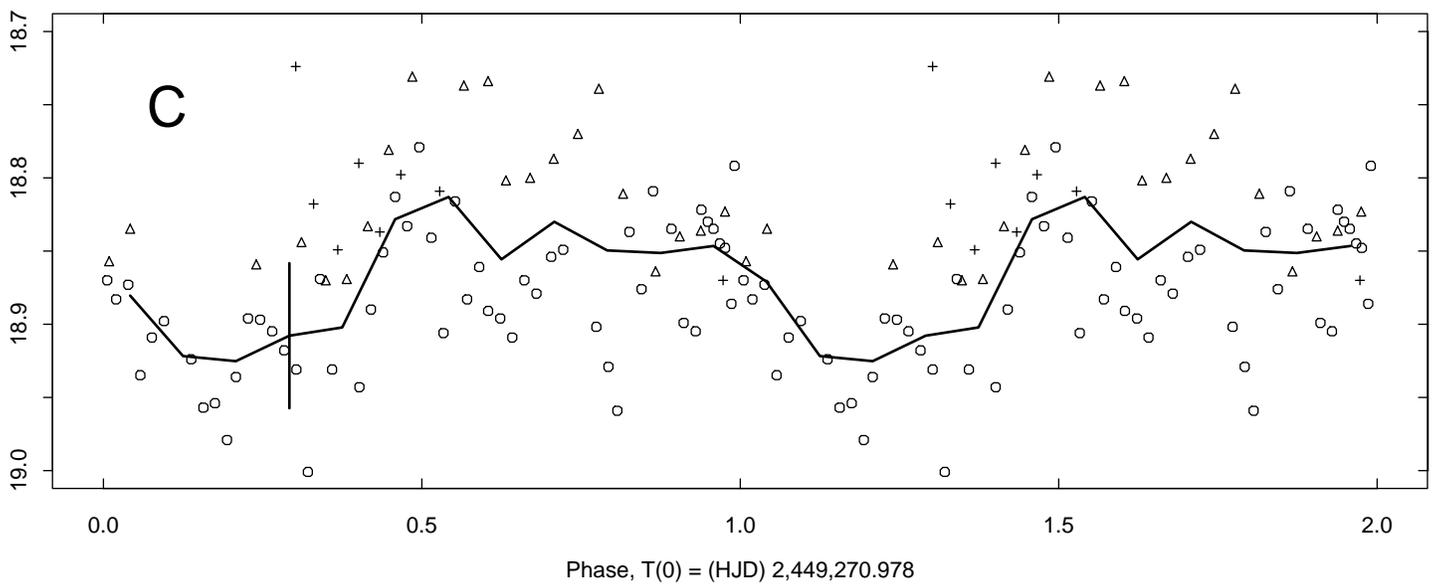
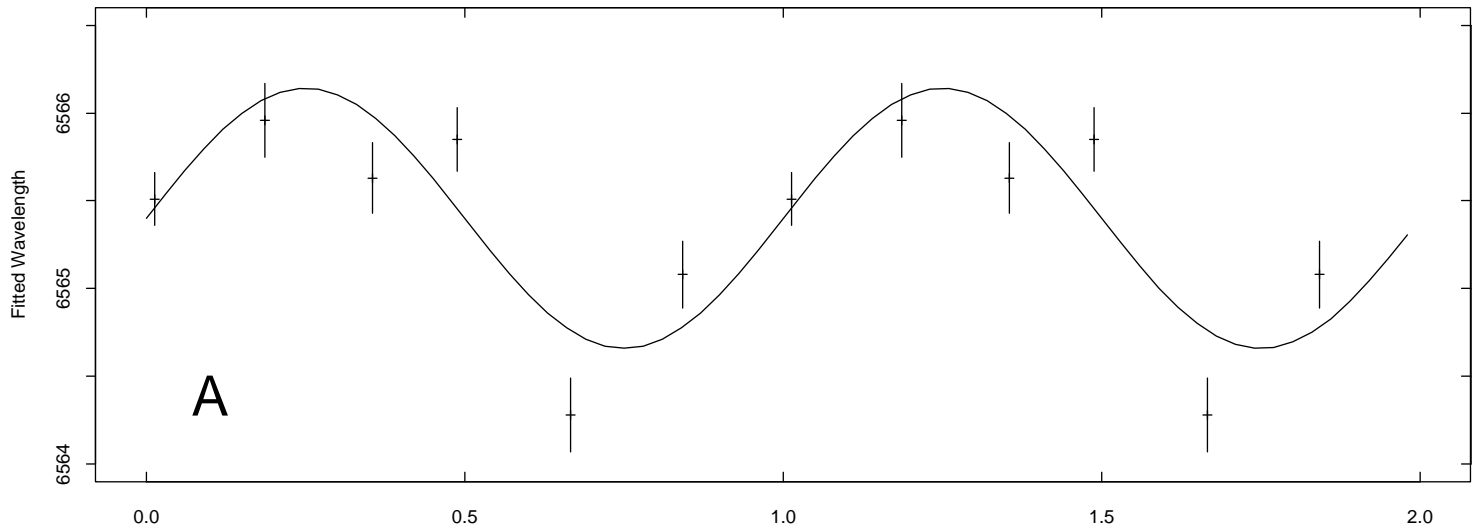
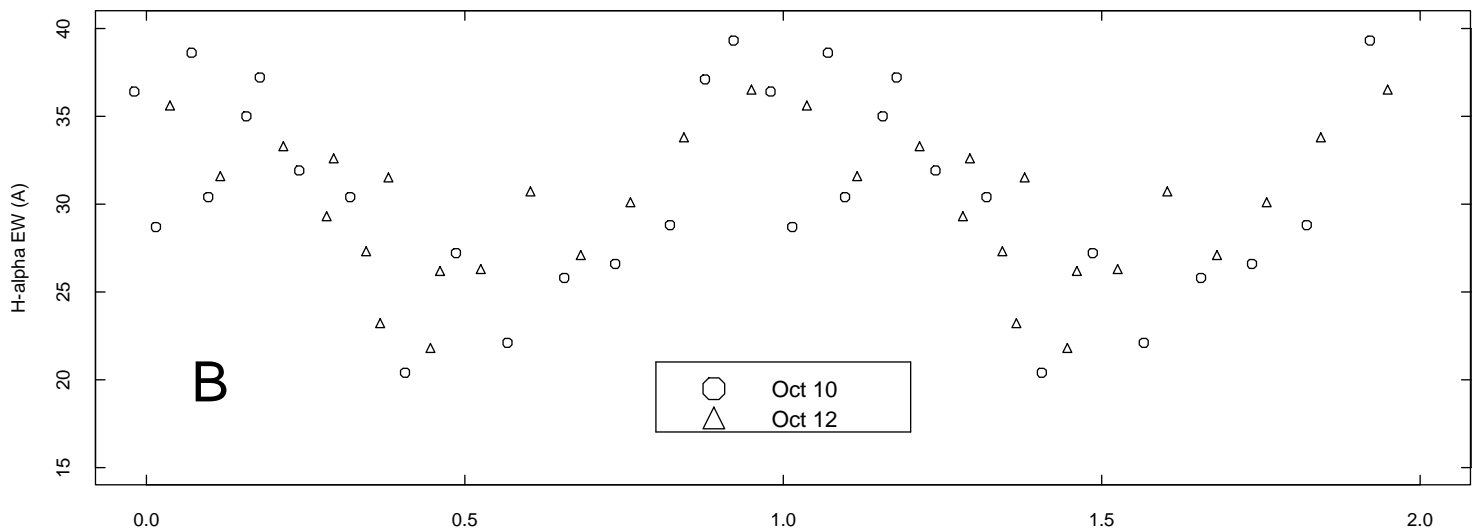
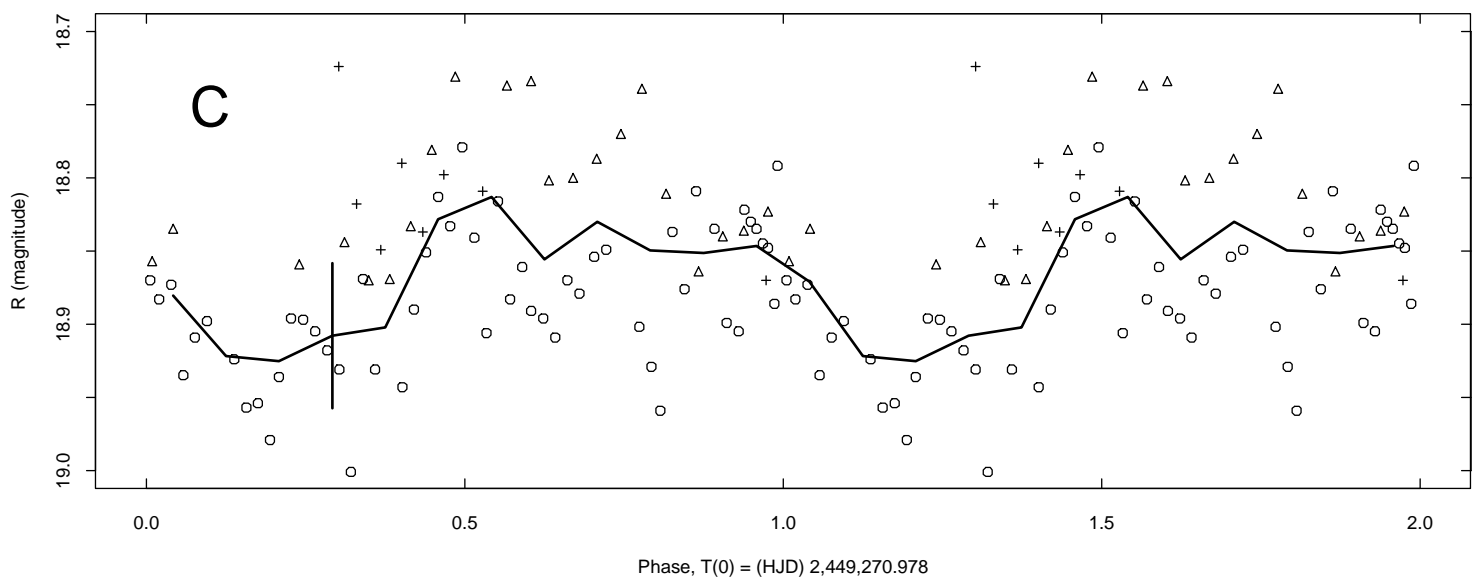
\end{document}